\documentclass[aps,prl,reprint,superscriptaddress]{revtex4-1}

\usepackage{amsmath,amssymb,amsfonts,bbm}
\usepackage[english]{babel}
\usepackage{graphicx}
\usepackage{color}

\begin{document}

\title{Solution to the Landau-Zener problem via Susskind-Glogower operators}

\author{B.M. Rodr\'{\i}guez-Lara}
\affiliation{Centre for Quantum Technologies, National University of Singapore, 2 Science Drive 3, Singapore 117542.}

\author{D. Rodr\'{\i}guez-M\'endez}
\affiliation{INAOE, Coordinaci\'on de Optica, Apdo. Postal 51 y 216, 72000 Puebla, Pue., Mexico.}

\author{H. Moya-Cessa}
\affiliation{INAOE, Coordinaci\'on de Optica, Apdo. Postal 51 y 216, 72000 Puebla, Pue., Mexico.}

\begin{abstract}
We show that, by means of a right-unitary transformation, the fully quantized Landau-Zener Hamiltonian in the weak-coupling regime may be solved by using known solutions from the standard Landau-Zener problem. 
In the strong-coupling regime, where the rotating wave approximation is not valid, we show that the quantized Landau-Zener Hamiltonian may be diagonalized in the atomic basis by means of a unitary transformation; hence allowing numerical solutions for the few photons regime via truncation.
\end{abstract}

\maketitle

\section{Introduction}

The Landau-Zener (LZ) problem \cite{Landau1932p46,Zener1932p696,Majorana1932p43} consists of a two-level system whose parameters are varied so that an anti-crossing of energy levels occurs. 
The transition between two energy states at an avoided level crossing in two- and multi-level systems is one of the few exactly solvable problems of time-dependent quantum evolution \cite{Vitanov1996p4288,Shytov2004p052708,Volkov2004p4069,Volkov2005p907}. 
Dynamics in atomic, molecular and mesoscopic systems can be described by the LZ process; see, for example, references within  \cite{Vitanov1996p4288,Shytov2004p052708}.
Recently, experimental realizations of many-body generalizations of LZ dynamics have been shown with ultracold atoms \cite{Chen2011p61} and theoretically analyzed considering strongly correlated bosons under fast sweeps \cite{Kasztelan2011p155302}.
The use of an interacting BEC driven in a bichromatic optical lattice has also been proposed as a realization of many-body nonlinear LZ dynamics \cite{Witthaut2006p063609,Witthaut2011p013609}.

In particular, there exists an approximate solution for a non-interacting many-body generalization of the LZ problem, including coupling to a quantized field, suggesting that many-body LZ physics can be profoundly different from the single two-level system interacting with a classical field case \cite{Altland2008p063602}. 
Motivated by these results, we analyze the LZ problem of the one two-level system coupled to a quantized field. 
First, we introduce a quantized LZ model and propose a realization in circuit quantum electrodynamics (circuit-QED) where weak- and strong-coupling regimes can be obtained; we also discuss plausible realizations of a formal many-body generalization of the model. 
Next, we present a right-unitary-transformation scheme to diagonalize the proposed quantized LZ model in the field basis under weak-coupling and show that the exact time evolution of the system is directly related to solutions of the standard LZ problem. 
Finally, we show a parity-related-transformation scheme to diagonalize the proposed quantized LZ model in the two-level system basis under strong-coupling and show that the resulting infinite set of differential equations is amenable for numerical solutions for a few starting excitations in the field.

\section{Mathematical model and physical realization}

A two-level system, that is, a qubit, driven by a quantized field is described by the model Hamiltonian, 
\begin{eqnarray}
\hat{H}_{DF} =  \frac{ \hbar}{2} \omega_{0} \hat{\sigma}_{z} + \hbar \omega \hat{a}^{\dagger} \hat{a} + \hbar g (\hat{a} + \hat{a}^{\dagger}) \hat{\sigma_{x}}. \label{eq:DipoleField}
\end{eqnarray}
Where the Pauli matrices associated to the qubit are given by the operators $\hat{\sigma}_{j}$, with $j=z,\pm$, and the symbol $\hat{a}$ ($\hat{a}^{\dagger}$) is the creation (annihilation) operator of the quantized field. 
The qubit two-level transition and the field frequencies are given by $\omega_{0}$ and $\omega$, respectively, and the qubit-field coupling by the parameter $g$.
 
In the weak-coupling regime,  where the values of the qubit-field coupling are at least an order of magnitude less than the qubit transition frequency,  $g \lesssim 0.1 \omega_0$, the rotating wave approximation (RWA) is valid and the Jaynes-Cummings (JC) model \cite{Jaynes1963p89},  in units of $\hbar$, describes the system,
\begin{eqnarray}
\hat{H}_{JC} = \frac{\omega_{0}}{2} \hat{\sigma}_{z} + \omega \hat{a}^{\dagger} \hat{a} + \lambda \left( \hat{a}^{\dagger}
\hat{\sigma}_{-} + \hat{a} \hat{\sigma}_{+} \right). \label{eq:JC}
\end{eqnarray}
The excitation number $\hat{N}= \hat{a}^{\dagger} \hat{a} + \hat{\sigma}_{z}/2$ is conserved by the JC model, $[\hat{N},\hat{H}_{JC}]=0$. 
Hence in a frame defined by the conserved excitation number rotating at the field frequency, $\omega$, by considering a time independent coupling, $\lambda$, and a time dependent detuning between the qubit and the field frequency given by the original Landau-Zener (LZ) process, $\Delta = \omega_{0}-\omega = - \tau \omega_{0}$, it is possible to write a quantized LZ, or LZ-JC Hamiltonian, in units of $\hbar \omega_{0}$, 
\begin{eqnarray}
\hat{H} = -\tau~ \hat{\sigma}_{z} + g \left( \hat{a}^{\dagger}
\hat{\sigma}_{-} + \hat{a} \hat{\sigma}_{+} \right). \label{eq:QLZH}
\end{eqnarray}
Where a scaled time $\tau = v^2 t$ and according coupling $g= \lambda/v^2$ has been used.
This particular choice of parameters means that the driving field is detuned to the blue of the qubit transition, $\omega = (1 + \tau) \omega_{0} = (1 + v^2 t) \omega_{0}$; the positive constant $v^2$ is just the steepness of the linear frequency ramp.
Choosing red detuning instead of blue is equivalent to replace $\tau \rightarrow -\tau$ in Eq.(\ref{eq:QLZH}).
The crossing of the spectra is given at $\tau \rightarrow 0$ where resonance is reached as $\omega_{0}=\omega{f}$. Notice that the quantized field frequency may be kept constant while adequately detuning the qubit transition.

A superconducting qubit coupled to a strip line resonator \cite{Blais2004p062320,JClarke2008p1031,Fink2008p315} may be described by the quantized LZ Hamiltonian in Eq.(\ref{eq:QLZH}) when the transition frequency of the charge (flux) qubit is varied linearly in time by a driving charge (magnetic flux) leading to $\omega_{0} = (1 - \tau) \omega$.
Furthermore, a circuit-QED implementation may allow the strong coupling needed to go beyond the RWA \cite{Niemczyk2010p772}. For simplicity, our analysis of the strong coupling case will use the Hamiltonian, in units of $\hbar \omega$,
\begin{eqnarray}
\hat{H}_{c} =  \tilde{\tau}~ \hat{\sigma}_{z} + \hat{a}^{\dagger} \hat{a} + g (\hat{a} + \hat{a}^{\dagger}) \hat{\sigma_{x}}, \label{eq:LZCH}
\end{eqnarray}
where the transition frequency of the qubit is tuned by an external charge (magnetic field) to vary linearly in time as $\tilde{\tau} = u^2 t$; again, the positive constant $u^2$ is just the steepness of the driving ramp.
In this model, the blue (red) detunning between the field and the qubit is provided by $\tau < 1$ ($\tau > 1$), and resonant driving is obtained at $\tau=1$.

A BEC in an asymmetric double well trap \cite{Smerzi1997p4950}, where the depth of one of the wells is made to vary linearly in time, may be described by a nonlinear version of the standard two-level Landau-Zener process.
In the mean-field approximation, classical dynamics of nonlinear LZ tunneling in Bose-Einstein condensates (BEC) has been studied and separated tunneling defined by the fixed points of the system found \cite{Itin2007p026218}.
A two-modes BEC driven by a quantized microwave field delivers a quantum version of the condensate in an asymmetric double-well model \cite{Chen2007p40004,RodriguezLara2011p016225}; the modes may be defined by two hyperfine structures of the ground state of a given atomic species, the time dependent detuning is given by the detuning between the qubit transition frequency and the driving field, and the nonlinearity may be tuned down by Feschbach resonances and neglected.
Such considerations deliver what we will call a Landau-Zener-Dicke (LZD) Hamiltonian,
\begin{eqnarray}
\hat{H}_{LZD} = -\tau~ \hat{S}_{z} +  g N_{q}^{-1/2} \left( \hat{a}^{\dagger}
\hat{S}_{-} + \hat{a} \hat{S}_{+} \right).\label{eq:QLZD}
\end{eqnarray}
Where the angular momentum basis related to angular momentum operators $\hat{S}_{j}$, with $j=z,\pm$, describes the ensemble containing $N_{q}$ qubits.
Also, the model in Eq.(\ref{eq:QLZD}) may describe the effective motion of a laser driven condensate, under the two-mode approximation, coupled to an optical cavity \cite{Nagy2008p127,Baumann2010p1301,Nagy2010p130401}, when the detuning between the laser pump and cavity frequencies with respect to the difference between the energies of the two-lowest-momentum modes is set to vary linearly in time.
While an approximate solution for the LZ problem is known for the  LZD model \cite{Altland2008p063602}, an exact solution for its time evolution is feasible but, at least for the time being, we restrict our analysis to the models of the one qubit within and without the rotating wave approximation described by Eq(\ref{eq:QLZH}) and Eq(\ref{eq:LZCH}), respectively.

\section{Diagonalization in the field basis}

In order to provide a time evolution operator for the quantum Landau-Zener Hamiltonian within the rotating wave approximation, Eq.(\ref{eq:QLZH}), we diagonalize it in the field basis by using the Susskind-Glogower \cite{Susskind1964p49} operators, 
\begin{eqnarray}
\hat{V} &=& \left( \hat{a}^{\dagger}\hat{a} +1  \right)^{-1/2}
\hat{a},
\end{eqnarray}
which are right-unitary, that is, $\hat{V} \hat{V}^{\dagger} = 1$ and $\hat{V}^{\dagger} \hat{V} = 1 - \vert 0\rangle\langle0\vert$. Their action on a Fock state of the number basis is given by $\hat{V} \vert n \rangle = \vert n-1 \rangle$ and $\hat{V}^{\dagger} \vert n \rangle = \vert n+1 \rangle$.

Via the Susskind-Glogower operators, the LZ-JC Hamiltonian in Eq.(\ref{eq:QLZH}) may be re-written as, 
\begin{eqnarray} \label{eq:RWQLZ}
\hat{H}=\tau \hat{H}_{0} + \hat{T}^{\dagger} \hat{H}_{LZ} \hat{T},
\end{eqnarray}
where the auxiliary, $\hat{H}_{0}$, and LZ-like, $\hat{H}_{LZ}$, Hamiltonians are
\begin{eqnarray}
\hat{H}_{0} &=& \left( {\bf I} - \hat{\sigma}_{z} \right) \vert 0\rangle\langle 0 \vert /2 \\
\hat{H}_{LZ} &=&  -\tau~ \hat{\sigma}_{z} + g \left(\hat{a}^{\dagger}\hat{a} +1  \right)^{1/2} \sigma_{x}. \label{eq:LZH} 
\end{eqnarray}
and the right-unitary transformation $\hat{T}$ is defined by \cite{Tang1996p154,JuarezAmaro2008p344,Klimov2009}
\begin{eqnarray} \label{eq:NUT}
\hat{T} &=& \left[ {\bf I} - \hat{\sigma}_{z} +\left( {\bf I} - \hat{\sigma}_{z}\right) \hat{V} \right] /2.
\end{eqnarray}
Notice that no approximation has been made in re-writing Eq.(\ref{eq:QLZH}) as Eq.(\ref{eq:RWQLZ}).
If a semi-classical quantization of the field were followed, like that proposed in Ref.\cite{Klimov1995p145} for time independent JC and Dicke models and only valid for coherent staes of the field, an approximate model lacking the $\hat{H}_0$ term would be obtained. 

As the auxiliary Hamiltonian $\hat{H}_{0}$ commutes with the transformed LZ-like Hamiltonian, $\hat{T}^{\dagger} \hat{H}_{LZ} \hat{T}$, at any given time, $\left[ \tau_{1} \hat{H}_{0},\hat{T}^{\dagger} \hat{H}_{LZ}(\tau_{2}) \hat{T} \right] = 0$, and using the fact that $\left[\hat{T}^{\dagger} \hat{H}_{LZ}(\tau_{1}) \hat{T} ,\hat{T}^{\dagger} \hat{H}_{LZ}(\tau_{2}) \hat{T} \right] = \hat{T}^{\dagger}  \left[\hat{H}_{LZ}(\tau_{1}),\hat{H}_{LZ}(\tau_{2}) \right] \hat{T} $, it is possible to write the time evolution operator of the system described by Eq.(\ref{eq:QLZH}) as
\begin{eqnarray}
\hat{U}(\tau) = \hat{U}_{0}(\tau) \hat{T}^{\dagger} \hat{U}_{LZ}(\tau) \hat{T};  \label{eq:QLZEv}
\end{eqnarray}
where it is trivial to find $\hat{U}_{0}(\tau) = e^{-i \tau \int\hat{H}_{0} dt} = e^{ - i \tau^2 H_{0} / 2}$ due to the fact that $\left[ \tau_{1} \hat{H}_{0}, \tau_{2} \hat{H}_{0}  \right] = 0$.

Notice that the right-unitary transformation $\hat{T}$ acting on the dressed state basis yields, 
\begin{eqnarray}
\hat{T} \left( c_{n,0} \vert n+1, 0 \rangle \pm c_{n,1} \vert n, 1 \rangle \right) &=& \vert n \rangle \left( c_{n,0} \vert 0 \rangle \pm c_{n,1} \vert 1 \rangle \right),  \nonumber \\
\end{eqnarray}
where the shorthand notation $\vert n, x \rangle \equiv \vert n \rangle \vert x \rangle \equiv \vert n \rangle_{field} \otimes \vert x \rangle_{atom}$ with $n=0,1,2,\ldots$ and $x =0,1$, has been used.
The time evolution operator for the LZ-like process, $\hat{U}_{LZ}(\tau)$, in the dressed state basis is given by the following set of coupled differential equations, 
\begin{eqnarray}
i~ \partial_{\tau} c_{n,1}(\tau) + \tau c_{n,1}(\tau) - g \left( n + 1\right)^{1/2} c_{n,0}(\tau) &=& 0,\nonumber \\
i ~\partial_{\tau} c_{n,0}(\tau) - \tau c_{n,0}(\tau) - g \left(n + 1\right)^{1/2} c_{n,1}(\tau) &=& 0, \nonumber \\
\end{eqnarray}
where the shorthand notation $\partial_{\tau} \cdot$ denotes the partial derivative with respect to the scaled time $\tau$.
Notice that this differential system is equivalent to that given by the standard two-level Landau-Zener process \cite{Vitanov1996p4288}, with the difference that the coupling between the two levels, $g$, is enhanced by a factor $\left( n + 1\right)^{1/2}$ due to the quantized field.

The system of coupled differential equations separates into, 
\begin{eqnarray} \label{eq:WDE}
\left[ \partial_{\tau} ^2 + \tau^2 + g^2 \left( n + 1 \right) - i \right] c_{n,1}(\tau) &=& 0, \nonumber \\
\left[ \partial_{\tau} ^2 + \tau^2 + g^2 \left( n + 1 \right) + i \right] c_{n,0}(\tau) &=& 0, 
\end{eqnarray}
which might be reduced to well-known differential equations accepting Whittaker functions, parabolic cylinder functions, and confluent  hypergeometric functions of the first kind as solutions \cite{Whittaker1927,Lebedev1965,Abramowitz1970,Prudnikov2003v3}. 
The latter will be preferred for the sake of simplicity at $t=0$; that is, even solutions evaluate to a non-zero constant while odd evaluate to zero. 
The properties of the hypergeometric function of the first kind, $_1F_1(\cdot,\cdot,\cdot)$, yield the time evolution for the amplitudes, up to a normalization and initial conditions factor,
\begin{eqnarray}
c_{1,e}(\tau)&=& e^{- i \tau^2/2} ~_1F_1\left(\frac{1}{2} + \frac{ i g_{n}^2}{4}, \frac{1}{2},  i \tau^2 \right),  \nonumber \\
c_{1,o}(\tau)&=&  -i g_{n} ~\tau ~e^{-i \tau^2/2} ~_1F_1\left(1 +
\frac{ i g_{n}^2}{4}, \frac{3}{2},   i \tau^2 \right), \nonumber \\
c_{0,e}(\tau)&=& e^{- i \tau^2/2}  ~_1F_1\left( \frac{i g_{n}^2}{4}, \frac{1}{2},  i \tau^2 \right), \nonumber \\
c_{0,o}(\tau)&=& - i  g_{n} ~\tau ~e^{-i \tau^2/2} ~_1F_1\left(\frac{1}{2} + \frac{ i g_{n}^2}{4}, \frac{3}{2}, i \tau^2 \right), \nonumber \\
\end{eqnarray}
where the functions imply time and photon number dependence, $c_{x,p}(\tau) \equiv c_{x,p}(g,n,\tau)$ with $x=0,1$ and $p=e,o$, the auxiliary characteristic values are defined as $g_{n} =g (n+1)^{1/2}$. The time evolution operator $\hat{U}_{LZ}$ is defined by the matrix elements, 
\begin{eqnarray}
U_{LZ}^{(i,j)} = u_{i,j} / \gamma, \quad n \rightarrow \hat{n}\label{eq:TELZ}
\end{eqnarray}
with
\begin{eqnarray}
u_{1,1}&=& c_{0,e}(\tau_{0}) c_{1,e}(\tau) - c_{0,o}(\tau_{0}) c_{1,o}(\tau), \nonumber \\
u_{1,2}&=& c_{1,e}(\tau_{0}) c_{1,o}(\tau) - c_{1,o}(\tau_{0}) c_{1,e}(\tau), \nonumber \\
u_{2,1}&=& c_{0,e}(\tau_{0}) c_{0,o}(\tau) - c_{0,o}(\tau_{0}) c_{0,e}(\tau), \nonumber \\
u_{2,2}&=& c_{1,e}(\tau_{0}) c_{0,e}(\tau) - c_{1,o}(\tau_{0}) c_{0,o}(\tau), \nonumber \\
\end{eqnarray}
where $u_{i,j} \equiv u_{i,j}(t, t_{0}, g_{\hat{n}})$ and the normalization factor is given by
\begin{eqnarray}
 \gamma &=& \left[ c_{0,e}(\tau_{0}) c_{1,e}(\tau_{0})- c_{0,o}(\tau_{0}) c_{1,o}(\tau_{0})  \right],
 \end{eqnarray}
with the number operator defined by $\hat{n} = \hat{a}^{\dagger}\hat{a}$. 
The time evolution operator $\hat{U}_{LZ}$ is exact, no approximation has been done.

As mentioned before, the presence of the quantized field enhances the qubit-field interaction simulating a standard two-level LZ process with effective coupling $g_{\hat{n}}$; this can be observed graphically from the time evolution of the population difference, 
\begin{eqnarray} 
\langle  \sigma_{z}(\tau) \rangle &=& \langle \psi(\tau_{0}) \vert \hat{U}^{\dagger}(\tau) \sigma_{z} \hat{U}(\tau)\psi(\tau_{0}) \rangle, \nonumber \\
&=& \langle \psi(\tau_{0}) \vert \hat{T}^{\dagger} \hat{U}_{LZ}^{\dagger}(\tau) \sigma_{z} \hat{U}_{LZ}(\tau) \hat{T} \vert \psi(\tau_{0}) \rangle \nonumber \\  \label{eq:TEPD}
\end{eqnarray} 

\begin{figure}
\includegraphics[width= 3.0in]{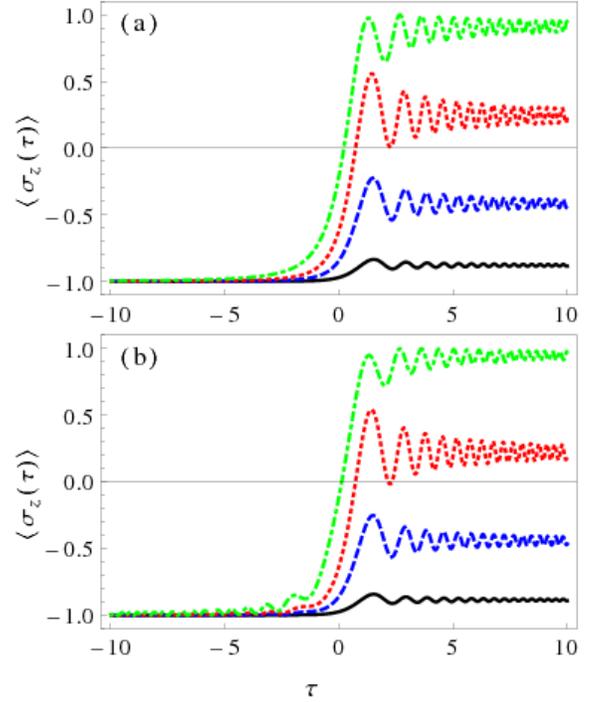}
\caption{(Color online) Exact evolution of the population difference $\langle \sigma_{z} \rangle$, Eq.(\ref{eq:TEPD}), for interactions starting at time (a) $\tau_{0} \rightarrow - \infty$ and (b) $\tau_{0} = -10$ under weak coupling $g = 0.1 \omega_{0}$ and initial state $\vert \psi_{0} \rangle = \vert n, g>$ with $n=1$ (solid black), $n=11$ (dashed blue), $n=31$ (dotted red) and $n=101$ (dot-dashed green).} \label{fig:Fig1}
\end{figure}

For the sake of historical comparison, Fig.\ref{fig:Fig1}(a) shows the time evolution of the population difference, $\langle  \sigma_{z}(\tau) \rangle$, in the context of the original LZ process where the interaction starts at $\tau_{i} \rightarrow - \infty$. Initial states given by $\vert \psi(\tau_{i} \rightarrow -\infty) \rangle= \vert n, g \rangle$ with $n \in \{\vert 1 \rangle,\vert 11 \rangle,\vert 31 \rangle,\vert 101 \rangle\}$ are considered. The results are exact numerics from Eq.(\ref{eq:TEPD}) by using Eq.(\ref{eq:TELZ}). 
Figure \ref{fig:Fig1}(b) shows the finite time effect described by the exact time evolution. 
Initial conditions are the same described above, but for an initial time of interaction $\tau_{0}=-10$.
As expected \cite{Vitanov1996p4288}, the population difference oscillates as soon as the finite time interaction starts.

\begin{figure}
\includegraphics[width=3.0in]{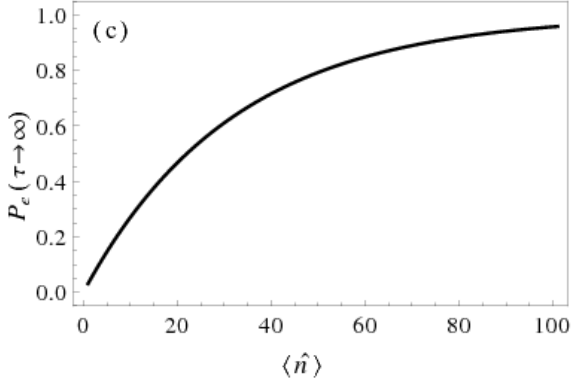}
\caption{(Color online) Asymptotic probability to find the qubit in the excited state, $P_{e}(\tau\rightarrow\infty)$, for the symmetric crossing defined by a starting (ending) time $\tau_{0} \rightarrow - \infty$ ($\tau \rightarrow \infty$) and the asymmetric crossing defined by $\tau_{0} \rightarrow -10$ ($\tau \rightarrow \infty$) under weak coupling $g = 0.1 \omega_{0}$ and initial state $\vert \psi_{0} \rangle = \vert n, g>$. Plots are calculated both with the analytical asymptotic expansion given by Eq.(\ref{eq:APE1}) and from numerics of the exact evolution, Eq.(\ref{eq:TEPD}) by substituting the hypergeometric function by its asymptotic expansion up to third order, with initial times $\tau_{0} \rightarrow -10^6$ and $\tau_{0} \rightarrow -10$ and final time $\tau \rightarrow 10^6$ (Plots overlap).} \label{fig:Fig2}
\end{figure}

Figure \ref{fig:Fig2} shows the asymptotic behavior for the excited state probability, $P_{e}$, for the symmetric crossing with starting (ending) times $\tau_{0} \rightarrow -\infty$ ($\tau \rightarrow \infty$) and the asymmetric crossing given by $\tau_{0} \rightarrow -10$ ($\tau \rightarrow \infty$). 
The qubit is initially taken in the ground state and the field at an arbitrary Fock state. 
The probabilities are calculated from the asymptotic expansion of Eq.(\ref{eq:TEPD}), via the parity and asymptotic properties of the hypergeometric function \cite{Lebedev1965}, and both yield the expression
\begin{eqnarray}
P_{e}(\tau\rightarrow \infty) &=& (1 + \langle \sigma_{z}(\tau\rightarrow\infty)\rangle)/2, \nonumber \\
& \approx& 1- e^{- \pi g_{n}^2},  \label{eq:APE1}
\end{eqnarray}
which is equivalent to the tunneling probability predicted in the standard LZ process.
The tunneling probability is enhanced by the number of photons in the quantized fied as $g_{n} = g (n+1)^{1/2}$.

\begin{figure}
\includegraphics[width= 3.0in]{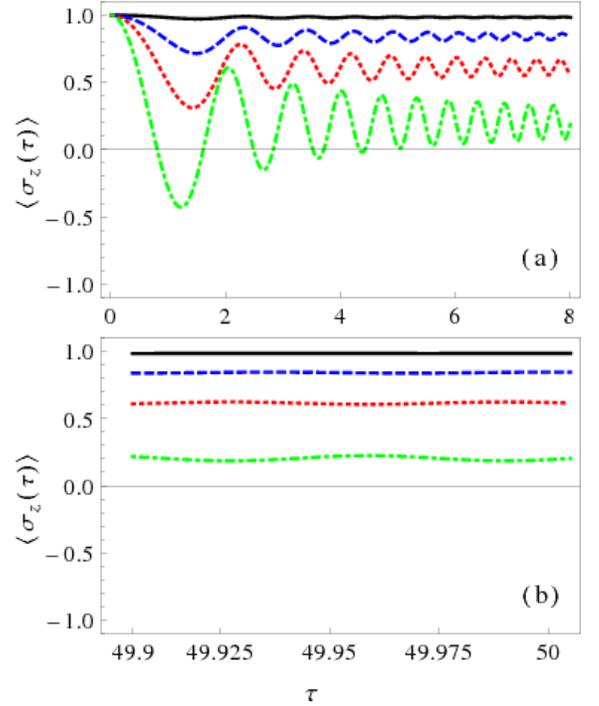}
\caption{(Color online) Exact evolution of the population difference $\langle \sigma_{z} \rangle$, Eq.(\ref{eq:TEPD}), for interactions starting at time $\tau_{0} = 0$ under weak coupling $g = 0.1 \omega_{0}$ and initial state $\vert \psi_{0} \rangle = \vert n, e>$ with $n=0$ (solid black), $n=10$ (dashed blue), $n=30$ (dotted red) and $n=100$ (dot-dashed green) for (a) short and (b) long interaction times} \label{fig:Fig3}
\end{figure}

\begin{figure}
\includegraphics[width=3.0in]{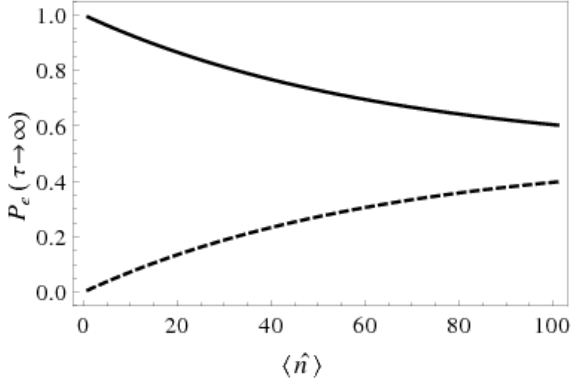}
\caption{(Color online) Asymptotic probability to find the qubit in the excited state, $P_{e}(\tau\rightarrow\infty)$, for the interaction defined in Fig.\ref{fig:Fig3} starting from initial states given by  $\vert \psi_{0} \rangle = \vert n,e \rangle$ (solid line) and $\vert \psi_{0} \rangle = \vert n,g \rangle$ (dashed line). Plots are calculated both with the analytical asymptotic expansion given by Eq.(\ref{eq:APE2}) and from numerics of the exact evolution, Eq.(\ref{eq:TEPD}) by substituting the hypergeometric function by its asymptotic expansion up to third order, with $\tau \rightarrow 10^6$ (Plots overlap). } \label{fig:Fig4}
\end{figure}

The exact time evolution operator found in this section describes any given set of parameters involving initial system state, weak coupling, initial and final time. 
Figure \ref{fig:Fig3} shows the time evolution of the population difference, $\langle  \sigma_{z}(\tau) \rangle$, in a purely asymmetric case starting from the crossing. The qubit is initialized in the excited state $\vert e \rangle$ and the field in the Fock states $n \in \{0,10,30,100\}]$; short and long interaction times are shown in subfigures Fig.\ref{fig:Fig3}(a) and Fig.\ref{fig:Fig3}(b), respectively.
In the asymptotic infinite time, the probability of finding the qubit in the excited state is
\begin{eqnarray}
P_{e}(\tau_{i}=0,\tau_{f} \rightarrow \infty)\approx \left(1 \pm e^{-\pi g_{n}^2/2} \right)/2, \label{eq:APE2}
\end{eqnarray}
depending if the qubit starts in the excited (plus sign) or the ground state (minus sing). This probability is is plotted in Fig.\ref{fig:Fig4} for the qubit starting in both the excited and the ground state.

\section{Diagonalization in the atomic basis}

For moderate and strong coupling, $g > 0.1 \omega_{0}$, the rotating wave approximation is not valid and the \textit{complete} Hamiltonian in Eq.(\ref{eq:LZCH}) has to be considered. Via the unitary transformation,
\begin{eqnarray}
\hat{R} = e^{-i \pi (\hat{\sigma}_{x} - 1) \hat{a}^{\dagger} \hat{a} /2},
\label{eq:RPar}
\end{eqnarray}
the complete Hamiltonian in Eq.(\ref{eq:LZCH}) becomes diagonal in the qubit basis,
\begin{eqnarray}
\hat{H}_R&=&\hat{R}^{\dagger} \hat{H}_C \hat{R} \nonumber \\
& =&  \tau~ \hat{\sigma}_{z} \cos(\pi \hat{a}^{\dagger} \hat{a}) + \hat{a}^{\dagger} \hat{a} + g (\hat{a} + \hat{a}^{\dagger}); \label{eq:DQB}
\end{eqnarray}
The use of the unitary transformation in Eq.(\ref{eq:RPar}) is equivalent to consider a \textit{parity chain} basis and the corresponding creation/annihilation operators as proposed in Ref.\cite{Casanova2010p263603} for the time independent qubit-field interaction in the strong-coupling regime.
It is straightforward to see the relation between these two approaches from
\begin{eqnarray}
\hat{R}^{\dagger} \hat{a} \hat{R} &=& \hat{a} e^{i \pi (\hat{\sigma}_{x} -1)/2} \nonumber \\ &=& \hat{a} \hat{\sigma}_{x}, \\
\hat{R}^{\dagger} \hat{a}^{\dagger} \hat{R} &=& \hat{a}^{\dagger} e^{-i \pi (\hat{\sigma_{x}} -1)/2} \nonumber \\ &=& \hat{a}^{\dagger} \hat{\sigma}_{x}, \\
\hat{R}^{\dagger} \hat{\sigma}_{z} \hat{R} &=& \hat{\sigma}_{z} \cos(\pi \hat{a}^{\dagger} \hat{a}) +
i \hat{\sigma}_{y} \sin(\pi \hat{a}^{\dagger} \hat{a}) \nonumber \\ &=&
\hat{\sigma}_{z} \cos(\pi \hat{a}^{\dagger} \hat{a}) \nonumber \\ &=&
(-1)^{\hat{a}^{\dagger} \hat{a}} \hat{\sigma}_{z},
\end{eqnarray}
as the parity operator, $\Pi = (-1)^{\hat{a}^{\dagger}\hat{a}} \sigma_{z}$, in the basis defined by the unitary Eq.(\ref{eq:RPar}) is given by
\begin{eqnarray}
\hat{R}^{\dagger} \Pi \hat{R} &=& (-1)^{\hat{a}^{\dagger}\hat{a}} \hat{R}^{\dagger} \sigma_{z} \hat{R}, \nonumber \\
&=& \sigma_{z}.
\end{eqnarray}
That is, the complete Hamiltonian Eq.(\ref{eq:DQB}) conserves parity, $[\hat{R}^{\dagger} \Pi \hat{R}, \hat{R}^{\dagger} \hat{H}_C \hat{R}] = [\hat{\sigma}_{z},\hat{H}_{R}]=0$.

\begin{figure}
\includegraphics[width= 3.0in]{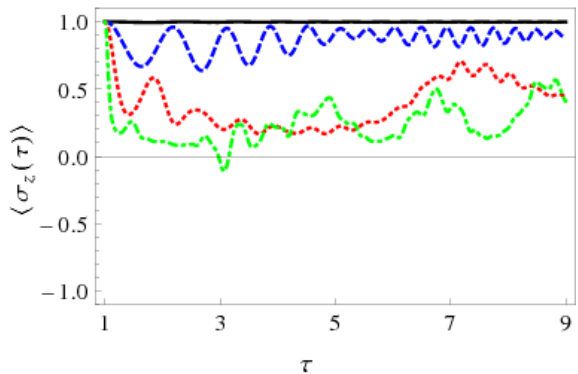}
\caption{(Color online) Numerical time evolution of the population difference $\langle \sigma_{z} \rangle$, Eq.(\ref{eq:TEPD2}), for interactions starting at time $\tau_{0} = 1$ under couplings $g = 0.1 \omega$ (solid black), $g=1 \omega$ (dashed blue), $g=3 \omega$ (dotted red) and $g=10 \omega$ (dot-dashed green) and initial state $\vert \psi_{0} \rangle = \vert 0, e>$. For the sake of comparison, the solid black line in this figure corresponds to initial conditions and parameters identical to the solid black line in Fig.\ref{fig:Fig3}. } \label{fig:Fig5}
\end{figure}

By moving into the rotating frame defined by the free field, $U_{F}(\tau) = e^{- i a^{\dagger} a \tau}$, the dynamics are given by the Hamiltonian, $H_{RF} = H_{R0} + H_{RI}$, with
\begin{eqnarray}
H_{R0} &=& \tau~ \hat{\sigma}_{z} \cos(\pi \hat{a}^{\dagger} \hat{a}) \nonumber \\
H_{RI} &=&  g (\hat{a} e^{i \tau} + \hat{a}^{\dagger} e^{-i \tau}).
\end{eqnarray}
The first of these terms,$H_{R0}$, is diagonal in both the qubit and Fock basis and commutes with itself at different scaled times, $[ H_{R0}(\tau_{1}),H_{R0}(\tau_{2}) ] = 0$; that is, it is possible to use a unitary transformation, 
\begin{eqnarray} \label{eq:UT}
U_{0}(t) = e^{- i \tau^2 \hat{\sigma}_{z} \cos(\pi \hat{a}^{\dagger} \hat{a}) /2 },
\end{eqnarray}
such that the system is described by,
\begin{eqnarray}
H_{RFU} =  g (\hat{a} e^{i \tau} + \hat{a}^{\dagger} e^{- i \tau}) e^{- i \tau^2 \hat{\sigma}_{z} \cos(\pi \hat{a}^{\dagger} \hat{a}) }.
\end{eqnarray}
This time dependent  Hamiltonian produces two infinite sets of coupled first order differential equations for the field, one for each qubit state  $x \in \{0,1\}$,
\begin{eqnarray}
i ~\partial_{\tau} c_{x,0}(\tau) &=& g e^{-i \tau} e^{\mp i \tau^2} c_{x,1}(\tau), 
\end{eqnarray}
for $n=0$ and
\begin{eqnarray}
i ~\partial_{\tau} c_{x,n}(\tau) &=& g n^{1/2} e^{i \tau} e^{\pm i \tau^2 (-1)^{n-1}} c_{x,n-1}(\tau) \nonumber \\
&& + g (n+1)^{1/2} e^{-i \tau} e^{\pm i \tau^2 (-1)^{n+1}} c_{x,n+1}(\tau) \nonumber \\ \label{eq:DES}
\end{eqnarray}
for $n\ge1$; for the sake of simplicity, dimension has been set to units $\hbar \omega$. The notation $ | \phi \rangle = \hat{R} | \psi \rangle = \sum_{x,n} c_{x,n}(\tau) |x,n\rangle$ has been used.
For Fock states with photon number $m$, the differential set defined by Eq.(\ref{eq:DES}) may be truncated at an arbitrary large $\tilde{n} \gg m$, 
\begin{eqnarray}
i ~\partial_{\tau} c_{x,\tilde{n}}(\tau) &=& n^{1/2} e^{i \tau} e^{\pm i \tau^2 (-1)^{\tilde{n}-1}} c_{x,\tilde{n}-1}(\tau)  \label{eq:DESTrunc}.
\end{eqnarray}
This is particularly helpful for initial states with small number of photons, in these cases numerical solutions may be given. Figure \ref{fig:Fig5} shows numerics for the population difference,
\begin{eqnarray}
\langle \sigma_{z}(\tau) \rangle = \langle \psi (\tau) \vert \sigma_{z} \cos \left(\pi \hat{a}^{\dagger} \hat{a} \right) \vert \psi (\tau) \rangle. \label{eq:TEPD2}
\end{eqnarray}
An initial state $\vert \psi_{0}(\tau=1)\rangle = \vert 0,e \rangle$ is taken and the set of coupled differential equations is truncated at length one hundred, that is, $m=100$. Qubit-field couplings in the range $g \in \{0.1,1,3,10\} \omega$ are considered. 
The initial time $\tau=1$ is chosen to emulate the case pictured in Fig.\ref{fig:Fig3}; both systems start from resonance, $\omega_{0} = \omega$. 
The black solid line in Fig.\ref{fig:Fig3} and Fig.\ref{fig:Fig5} represents identical initial conditions, $\vert \psi_{0} \rangle = \vert 0,e \rangle$ and coupling $g=0.1 \omega$ and deliver similar dynamics. 
The rest of the couplings treated in Fig.\ref{fig:Fig5} show dynamics similar to those under the RWA, Eq.(\ref{eq:RWQLZ}), for small normalized times, $\tau \ll 1$ and, then, the coherent oscillations break due the action of the counter-rotating terms, as expected.

\section{Conclusion}

We have presented a right-unitary approach to solve the qubit-quantized-field interaction under the rotating wave approximation with frequency detuning varying linearly in time, Eq.(\ref{eq:QLZH}). 
The model may be realized in circuit-QED \cite{Blais2004p062320}.
We have diagonalized the model Hamiltonian, Eq.(\ref{eq:QLZH}), in the quantized field basis and shown that the procedure to obtain the non-trivial ingredient of the evolution, $\hat{U}_{LZ}(t)$ in Eq.(\ref{eq:QLZEv}), is already known from the interaction of a classical field with a qubit \cite{Vitanov1996p4288}.
The presented solution is exact.
Its analytical closed form allows its use in modular scenarios to engineer particular states or Hamiltonians; compare, for example, with Ref.\cite{Vitanov1996p4288}, where different symmetric and asymmetric crossings in the standard Landau-Zener model are proposed and may be used for state engineering, or Ref.\cite{Tsomokos2010p052311}, where modular cavity-QED is proposed to engineer exotic lattice systems.
In the asymptotic symmetric case equivalent to the standard LZ problem, the quantized version for initial separable states presents a similar asymptotic behavior for the probability of the LZ transition, $P_{e}(\tau\rightarrow \infty) \approx 1- e^{- \pi g_{n}^2}$ in Eq.(\ref{eq:APE1}), with the distinction of an enhanced coupling proportional to the square root of the number of photons in the initial state, $g_{\hat{n}} = g (\hat{n} + 1)^{1/2}$. 

The strong coupling dynamics of the system, where the rotating wave approximation is not valid, has been studied via an unitary transformation that diagonalizes the Hamiltonian Eq.(\ref{eq:LZCH}) in the qubit basis.
This operator approach is comparable to defining a \textit{parity chain} basis for the system \cite{Casanova2010p263603}.
The system dynamics is given by an infinite set of coupled differential equations amenable to numerical solutions, via truncation, for small initial number of excitations in the quantized field.

\section*{Acknowledgement}
B.M.R.L. is grateful to C. Noh for fruitful discussion.


\begin{thebibliography}{35}
\expandafter\ifx\csname natexlab\endcsname\relax\def\natexlab#1{#1}\fi
\expandafter\ifx\csname bibnamefont\endcsname\relax
  \def\bibnamefont#1{#1}\fi
\expandafter\ifx\csname bibfnamefont\endcsname\relax
  \def\bibfnamefont#1{#1}\fi
\expandafter\ifx\csname citenamefont\endcsname\relax
  \def\citenamefont#1{#1}\fi
\expandafter\ifx\csname url\endcsname\relax
  \def\url#1{\texttt{#1}}\fi
\expandafter\ifx\csname urlprefix\endcsname\relax\def\urlprefix{URL }\fi
\providecommand{\bibinfo}[2]{#2}
\providecommand{\eprint}[2][]{\url{#2}}

\bibitem[{\citenamefont{Landau}(1932)}]{Landau1932p46}
\bibinfo{author}{\bibfnamefont{L.}~\bibnamefont{Landau}},
  \bibinfo{journal}{Physik. Z. Sowjet} \textbf{\bibinfo{volume}{2}},
  \bibinfo{pages}{46 } (\bibinfo{year}{1932}).

\bibitem[{\citenamefont{Zener}(1932)}]{Zener1932p696}
\bibinfo{author}{\bibfnamefont{C.}~\bibnamefont{Zener}},
  \bibinfo{journal}{Proc. R. Soc. Lond. A} \textbf{\bibinfo{volume}{137}},
  \bibinfo{pages}{696 } (\bibinfo{year}{1932}).

\bibitem[{\citenamefont{Majorana}(1932)}]{Majorana1932p43}
\bibinfo{author}{\bibfnamefont{E.}~\bibnamefont{Majorana}},
  \bibinfo{journal}{Nuovo Cimento} \textbf{\bibinfo{volume}{9}},
  \bibinfo{pages}{43 } (\bibinfo{year}{1932}).

\bibitem[{\citenamefont{Vitanov and Garraway}(1996)}]{Vitanov1996p4288}
\bibinfo{author}{\bibfnamefont{N.~V.} \bibnamefont{Vitanov}} \bibnamefont{and}
  \bibinfo{author}{\bibfnamefont{B.~M.} \bibnamefont{Garraway}},
  \bibinfo{journal}{Phys. Rev. A} \textbf{\bibinfo{volume}{53}},
  \bibinfo{pages}{4288 } (\bibinfo{year}{1996}).

\bibitem[{\citenamefont{Shytov}(2004)}]{Shytov2004p052708}
\bibinfo{author}{\bibfnamefont{A.~V.} \bibnamefont{Shytov}},
  \bibinfo{journal}{Phys. Rev. A} \textbf{\bibinfo{volume}{70}},
  \bibinfo{pages}{052708} (\bibinfo{year}{2004}).

\bibitem[{\citenamefont{Volkov and Ostrovsky}(2004)}]{Volkov2004p4069}
\bibinfo{author}{\bibfnamefont{M.~V.} \bibnamefont{Volkov}} \bibnamefont{and}
  \bibinfo{author}{\bibfnamefont{V.~N.} \bibnamefont{Ostrovsky}},
  \bibinfo{journal}{J. Phys. B: At. Mol. Opt. Phys.}
  \textbf{\bibinfo{volume}{37}}, \bibinfo{pages}{4069 } (\bibinfo{year}{2004}).

\bibitem[{\citenamefont{Volkov and Ostrovsky}(2005)}]{Volkov2005p907}
\bibinfo{author}{\bibfnamefont{M.~V.} \bibnamefont{Volkov}} \bibnamefont{and}
  \bibinfo{author}{\bibfnamefont{V.~N.} \bibnamefont{Ostrovsky}},
  \bibinfo{journal}{J. Phys. B: At. Mol. Opt. Phys.}
  \textbf{\bibinfo{volume}{38}}, \bibinfo{pages}{907 } (\bibinfo{year}{2005}).

\bibitem[{\citenamefont{Chen et~al.}(2011)\citenamefont{Chen, Huber, Trotzky,
  Bloch, and Altman}}]{Chen2011p61}
\bibinfo{author}{\bibfnamefont{Y.-A.} \bibnamefont{Chen}},
  \bibinfo{author}{\bibfnamefont{S.~D.} \bibnamefont{Huber}},
  \bibinfo{author}{\bibfnamefont{S.}~\bibnamefont{Trotzky}},
  \bibinfo{author}{\bibfnamefont{I.}~\bibnamefont{Bloch}}, \bibnamefont{and}
  \bibinfo{author}{\bibfnamefont{E.}~\bibnamefont{Altman}},
  \bibinfo{journal}{Nature Physics} \textbf{\bibinfo{volume}{7}},
  \bibinfo{pages}{61 } (\bibinfo{year}{2011}).

\bibitem[{\citenamefont{Kasztelan et~al.}(2011)\citenamefont{Kasztelan,
  Trotzky, Chen, Bloch, McCulloch, Schollwock, and
  Orso}}]{Kasztelan2011p155302}
\bibinfo{author}{\bibfnamefont{C.}~\bibnamefont{Kasztelan}},
  \bibinfo{author}{\bibfnamefont{S.}~\bibnamefont{Trotzky}},
  \bibinfo{author}{\bibfnamefont{Y.-A.} \bibnamefont{Chen}},
  \bibinfo{author}{\bibfnamefont{I.}~\bibnamefont{Bloch}},
  \bibinfo{author}{\bibfnamefont{I.~P.} \bibnamefont{McCulloch}},
  \bibinfo{author}{\bibfnamefont{U.}~\bibnamefont{Schollwock}},
  \bibnamefont{and} \bibinfo{author}{\bibfnamefont{G.}~\bibnamefont{Orso}},
  \bibinfo{journal}{Phys. Rev. Lett.} \textbf{\bibinfo{volume}{106}},
  \bibinfo{pages}{155302} (\bibinfo{year}{2011}).

\bibitem[{\citenamefont{Witthaut et~al.}(2006)\citenamefont{Witthaut, Graefe,
  and Korsch}}]{Witthaut2006p063609}
\bibinfo{author}{\bibfnamefont{D.}~\bibnamefont{Witthaut}},
  \bibinfo{author}{\bibfnamefont{E.~M.} \bibnamefont{Graefe}},
  \bibnamefont{and} \bibinfo{author}{\bibfnamefont{H.~J.}
  \bibnamefont{Korsch}}, \bibinfo{journal}{Phys. Rev. A}
  \textbf{\bibinfo{volume}{73}}, \bibinfo{pages}{063609}
  (\bibinfo{year}{2006}).

\bibitem[{\citenamefont{Witthaut et~al.}(2011)\citenamefont{Witthaut, Trimborn,
  Kegel, and Korsch}}]{Witthaut2011p013609}
\bibinfo{author}{\bibfnamefont{D.}~\bibnamefont{Witthaut}},
  \bibinfo{author}{\bibfnamefont{F.}~\bibnamefont{Trimborn}},
  \bibinfo{author}{\bibfnamefont{V.}~\bibnamefont{Kegel}}, \bibnamefont{and}
  \bibinfo{author}{\bibfnamefont{H.~J.} \bibnamefont{Korsch}},
  \bibinfo{journal}{Phys. Rev. A} \textbf{\bibinfo{volume}{83}},
  \bibinfo{pages}{013609} (\bibinfo{year}{2011}).

\bibitem[{\citenamefont{Altland and Gurarie}(2008)}]{Altland2008p063602}
\bibinfo{author}{\bibfnamefont{A.}~\bibnamefont{Altland}} \bibnamefont{and}
  \bibinfo{author}{\bibfnamefont{V.}~\bibnamefont{Gurarie}},
  \bibinfo{journal}{Phys. Rev. Lett.} \textbf{\bibinfo{volume}{100}},
  \bibinfo{pages}{063602} (\bibinfo{year}{2008}).

\bibitem[{\citenamefont{Jaynes and Cummingss}(1963)}]{Jaynes1963p89}
\bibinfo{author}{\bibfnamefont{E.~T.} \bibnamefont{Jaynes}} \bibnamefont{and}
  \bibinfo{author}{\bibfnamefont{F.~W.} \bibnamefont{Cummingss}},
  \bibinfo{journal}{Proc. IEEE} \textbf{\bibinfo{volume}{51}},
  \bibinfo{pages}{89 } (\bibinfo{year}{1963}).

\bibitem[{\citenamefont{Blais et~al.}(2004)\citenamefont{Blais, Huang,
  Wallraff, Girvin, and Schoelkopf}}]{Blais2004p062320}
\bibinfo{author}{\bibfnamefont{A.}~\bibnamefont{Blais}},
  \bibinfo{author}{\bibfnamefont{R.-S.} \bibnamefont{Huang}},
  \bibinfo{author}{\bibfnamefont{A.}~\bibnamefont{Wallraff}},
  \bibinfo{author}{\bibfnamefont{S.~M.} \bibnamefont{Girvin}},
  \bibnamefont{and} \bibinfo{author}{\bibfnamefont{R.~J.}
  \bibnamefont{Schoelkopf}}, \bibinfo{journal}{Phys. Rev. A}
  \textbf{\bibinfo{volume}{69}}, \bibinfo{pages}{062320}
  (\bibinfo{year}{2004}).

\bibitem[{\citenamefont{Clarke and Wilhelm}(2008)}]{JClarke2008p1031}
\bibinfo{author}{\bibfnamefont{J.}~\bibnamefont{Clarke}} \bibnamefont{and}
  \bibinfo{author}{\bibfnamefont{F.~K.} \bibnamefont{Wilhelm}},
  \bibinfo{journal}{Nature} \textbf{\bibinfo{volume}{453}},
  \bibinfo{pages}{1031 } (\bibinfo{year}{2008}).

\bibitem[{\citenamefont{Fink et~al.}(2008)\citenamefont{Fink, Goppl, Baur,
  Bianchetti, Leek, Blais, and Wallraff}}]{Fink2008p315}
\bibinfo{author}{\bibfnamefont{J.~M.} \bibnamefont{Fink}},
  \bibinfo{author}{\bibfnamefont{M.}~\bibnamefont{Goppl}},
  \bibinfo{author}{\bibfnamefont{M.}~\bibnamefont{Baur}},
  \bibinfo{author}{\bibfnamefont{R.}~\bibnamefont{Bianchetti}},
  \bibinfo{author}{\bibfnamefont{P.~J.} \bibnamefont{Leek}},
  \bibinfo{author}{\bibfnamefont{A.}~\bibnamefont{Blais}}, \bibnamefont{and}
  \bibinfo{author}{\bibfnamefont{A.}~\bibnamefont{Wallraff}},
  \bibinfo{journal}{Nature} \textbf{\bibinfo{volume}{454}}, \bibinfo{pages}{315
  } (\bibinfo{year}{2008}).

\bibitem[{\citenamefont{Niemczyk et~al.}(2010)\citenamefont{Niemczyk, Deppe,
  Huebl, Menzel, Hocke, Schwarz, Garcia-Ripoll, Zueco, H{\"u}mmer, Solano
  et~al.}}]{Niemczyk2010p772}
\bibinfo{author}{\bibfnamefont{T.}~\bibnamefont{Niemczyk}},
  \bibinfo{author}{\bibfnamefont{F.}~\bibnamefont{Deppe}},
  \bibinfo{author}{\bibfnamefont{H.}~\bibnamefont{Huebl}},
  \bibinfo{author}{\bibfnamefont{E.~P.} \bibnamefont{Menzel}},
  \bibinfo{author}{\bibfnamefont{F.}~\bibnamefont{Hocke}},
  \bibinfo{author}{\bibfnamefont{M.~J.} \bibnamefont{Schwarz}},
  \bibinfo{author}{\bibfnamefont{J.~J.} \bibnamefont{Garcia-Ripoll}},
  \bibinfo{author}{\bibfnamefont{D.}~\bibnamefont{Zueco}},
  \bibinfo{author}{\bibfnamefont{T.}~\bibnamefont{H{\"u}mmer}},
  \bibinfo{author}{\bibfnamefont{E.}~\bibnamefont{Solano}},
  \bibnamefont{et~al.}, \bibinfo{journal}{Nature Phys.}
  \textbf{\bibinfo{volume}{6}}, \bibinfo{pages}{772 } (\bibinfo{year}{2010}).

\bibitem[{\citenamefont{Smerzi et~al.}(1997)\citenamefont{Smerzi, Fantoni,
  Giovanazzi, and Shenoy}}]{Smerzi1997p4950}
\bibinfo{author}{\bibfnamefont{A.}~\bibnamefont{Smerzi}},
  \bibinfo{author}{\bibfnamefont{S.}~\bibnamefont{Fantoni}},
  \bibinfo{author}{\bibfnamefont{S.}~\bibnamefont{Giovanazzi}},
  \bibnamefont{and} \bibinfo{author}{\bibfnamefont{S.~R.}
  \bibnamefont{Shenoy}}, \bibinfo{journal}{Phys. Rev. Lett.}
  \textbf{\bibinfo{volume}{79}}, \bibinfo{pages}{4950 } (\bibinfo{year}{1997}).

\bibitem[{\citenamefont{Itin and Watanabe}(2007)}]{Itin2007p026218}
\bibinfo{author}{\bibfnamefont{A.~P.} \bibnamefont{Itin}} \bibnamefont{and}
  \bibinfo{author}{\bibfnamefont{S.}~\bibnamefont{Watanabe}},
  \bibinfo{journal}{Phys. Rev. E} \textbf{\bibinfo{volume}{76}},
  \bibinfo{pages}{026218} (\bibinfo{year}{2007}).

\bibitem[{\citenamefont{Chen et~al.}(2007)\citenamefont{Chen, Chen, and
  Liang}}]{Chen2007p40004}
\bibinfo{author}{\bibfnamefont{G.}~\bibnamefont{Chen}},
  \bibinfo{author}{\bibfnamefont{Z.}~\bibnamefont{Chen}}, \bibnamefont{and}
  \bibinfo{author}{\bibfnamefont{J.-Q.} \bibnamefont{Liang}},
  \bibinfo{journal}{Europhys. Lett.} \textbf{\bibinfo{volume}{80}},
  \bibinfo{pages}{40004} (\bibinfo{year}{2007}).

\bibitem[{\citenamefont{Rodr{\'\i}iguez-Lara and
  Lee}(2011)}]{RodriguezLara2011p016225}
\bibinfo{author}{\bibfnamefont{B.~M.} \bibnamefont{Rodr{\'\i}iguez-Lara}}
  \bibnamefont{and} \bibinfo{author}{\bibfnamefont{R.-K.} \bibnamefont{Lee}},
  \bibinfo{journal}{Phys. Rev. E} \textbf{\bibinfo{volume}{84}},
  \bibinfo{pages}{016225} (\bibinfo{year}{2011}).

\bibitem[{\citenamefont{Nagy et~al.}(2008)\citenamefont{Nagy, Szirmai, and
  Domokos}}]{Nagy2008p127}
\bibinfo{author}{\bibfnamefont{D.}~\bibnamefont{Nagy}},
  \bibinfo{author}{\bibfnamefont{G.}~\bibnamefont{Szirmai}}, \bibnamefont{and}
  \bibinfo{author}{\bibfnamefont{P.}~\bibnamefont{Domokos}},
  \bibinfo{journal}{Eur. Phys. J. D} \textbf{\bibinfo{volume}{48}},
  \bibinfo{pages}{127 } (\bibinfo{year}{2008}).

\bibitem[{\citenamefont{Baumann et~al.}(2010)\citenamefont{Baumann, Guerlin,
  Brennecke, and Esslinger}}]{Baumann2010p1301}
\bibinfo{author}{\bibfnamefont{K.}~\bibnamefont{Baumann}},
  \bibinfo{author}{\bibfnamefont{C.}~\bibnamefont{Guerlin}},
  \bibinfo{author}{\bibfnamefont{F.}~\bibnamefont{Brennecke}},
  \bibnamefont{and}
  \bibinfo{author}{\bibfnamefont{T.}~\bibnamefont{Esslinger}},
  \bibinfo{journal}{Nature} \textbf{\bibinfo{volume}{464}},
  \bibinfo{pages}{1301 } (\bibinfo{year}{2010}).

\bibitem[{\citenamefont{Nagy et~al.}(2010)\citenamefont{Nagy, Konya, Szirmai,
  and Domokos}}]{Nagy2010p130401}
\bibinfo{author}{\bibfnamefont{D.}~\bibnamefont{Nagy}},
  \bibinfo{author}{\bibfnamefont{G.}~\bibnamefont{Konya}},
  \bibinfo{author}{\bibfnamefont{G.}~\bibnamefont{Szirmai}}, \bibnamefont{and}
  \bibinfo{author}{\bibfnamefont{P.}~\bibnamefont{Domokos}},
  \bibinfo{journal}{Phys. Rev. Lett.} \textbf{\bibinfo{volume}{104}},
  \bibinfo{pages}{130401} (\bibinfo{year}{2010}).

\bibitem[{\citenamefont{Susskind and Glogower}(1964)}]{Susskind1964p49}
\bibinfo{author}{\bibfnamefont{L.}~\bibnamefont{Susskind}} \bibnamefont{and}
  \bibinfo{author}{\bibfnamefont{J.}~\bibnamefont{Glogower}},
  \bibinfo{journal}{Physics} \textbf{\bibinfo{volume}{1}}, \bibinfo{pages}{49}
  (\bibinfo{year}{1964}).

\bibitem[{\citenamefont{Tang}(1996)}]{Tang1996p154}
\bibinfo{author}{\bibfnamefont{Z.}~\bibnamefont{Tang}}, \bibinfo{journal}{Phys.
  Rev. A} \textbf{\bibinfo{volume}{54}}, \bibinfo{pages}{154 }
  (\bibinfo{year}{1996}).

\bibitem[{\citenamefont{Ju{\'a}rez-Amaro
  et~al.}(2008)\citenamefont{Ju{\'a}rez-Amaro, Vargas-Mart{\'i}nez, and
  Moya-Cessa}}]{JuarezAmaro2008p344}
\bibinfo{author}{\bibfnamefont{R.}~\bibnamefont{Ju{\'a}rez-Amaro}},
  \bibinfo{author}{\bibfnamefont{J.~M.} \bibnamefont{Vargas-Mart{\'i}nez}},
  \bibnamefont{and}
  \bibinfo{author}{\bibfnamefont{H.}~\bibnamefont{Moya-Cessa}},
  \bibinfo{journal}{Laser Physics} \textbf{\bibinfo{volume}{18}},
  \bibinfo{pages}{344 } (\bibinfo{year}{2008}).

\bibitem[{\citenamefont{Klimov and Chumakov}(2009)}]{Klimov2009}
\bibinfo{author}{\bibfnamefont{A.~B.} \bibnamefont{Klimov}} \bibnamefont{and}
  \bibinfo{author}{\bibfnamefont{S.~M.} \bibnamefont{Chumakov}},
  \emph{\bibinfo{title}{A group theoretical approach to quantum optics}}
  (\bibinfo{publisher}{Wiley-VCH}, \bibinfo{year}{2009}).

\bibitem[{\citenamefont{Klimov and Chumakov}(1995)}]{Klimov1995p145}
\bibinfo{author}{\bibfnamefont{A.~B.} \bibnamefont{Klimov}} \bibnamefont{and}
  \bibinfo{author}{\bibfnamefont{S.~M.} \bibnamefont{Chumakov}},
  \bibinfo{journal}{Phys. Lett. A} \textbf{\bibinfo{volume}{202}},
  \bibinfo{pages}{145 } (\bibinfo{year}{1995}).

\bibitem[{\citenamefont{Whittaker and Watson}(1927)}]{Whittaker1927}
\bibinfo{author}{\bibfnamefont{E.~T.} \bibnamefont{Whittaker}}
  \bibnamefont{and} \bibinfo{author}{\bibfnamefont{G.~N.}
  \bibnamefont{Watson}}, \emph{\bibinfo{title}{A course of modern analysis}}
  (\bibinfo{publisher}{Cambridge University Press}, \bibinfo{year}{1927}).

\bibitem[{\citenamefont{Lebedev}(1965)}]{Lebedev1965}
\bibinfo{author}{\bibfnamefont{N.~N.} \bibnamefont{Lebedev}},
  \emph{\bibinfo{title}{Special functions and their applications}}
  (\bibinfo{publisher}{Prentice-Hall}, \bibinfo{year}{1965}).

\bibitem[{\citenamefont{Abramowitz and Stegun}(1970)}]{Abramowitz1970}
\bibinfo{author}{\bibfnamefont{M.}~\bibnamefont{Abramowitz}} \bibnamefont{and}
  \bibinfo{author}{\bibfnamefont{I.~A.} \bibnamefont{Stegun}},
  \emph{\bibinfo{title}{Handbook of Mathematical Functions}}
  (\bibinfo{year}{1970}).

\bibitem[{\citenamefont{Prudnikov et~al.}(2003)\citenamefont{Prudnikov,
  Brychkov, and Marichev}}]{Prudnikov2003v3}
\bibinfo{author}{\bibfnamefont{A.~P.} \bibnamefont{Prudnikov}},
  \bibinfo{author}{\bibfnamefont{J.~A.} \bibnamefont{Brychkov}},
  \bibnamefont{and} \bibinfo{author}{\bibfnamefont{O.~I.}
  \bibnamefont{Marichev}}, \emph{\bibinfo{title}{Integrals and series}},
  vol.~\bibinfo{volume}{3} (\bibinfo{publisher}{Mockba}, \bibinfo{year}{2003}).

\bibitem[{\citenamefont{Casanova et~al.}(2010)\citenamefont{Casanova, Romero,
  Lizuain, Garc\'ia-Ripoll, and Solano}}]{Casanova2010p263603}
\bibinfo{author}{\bibfnamefont{J.}~\bibnamefont{Casanova}},
  \bibinfo{author}{\bibfnamefont{G.}~\bibnamefont{Romero}},
  \bibinfo{author}{\bibfnamefont{I.}~\bibnamefont{Lizuain}},
  \bibinfo{author}{\bibfnamefont{J.~J.} \bibnamefont{Garc\'ia-Ripoll}},
  \bibnamefont{and} \bibinfo{author}{\bibfnamefont{E.}~\bibnamefont{Solano}},
  \bibinfo{journal}{Phys. Rev. Lett.} \textbf{\bibinfo{volume}{105}},
  \bibinfo{pages}{263603} (\bibinfo{year}{2010}).

\bibitem[{\citenamefont{Tsomokos et~al.}(2010)\citenamefont{Tsomokos, Ashhab,
  and Nori}}]{Tsomokos2010p052311}
\bibinfo{author}{\bibfnamefont{D.~I.} \bibnamefont{Tsomokos}},
  \bibinfo{author}{\bibfnamefont{S.}~\bibnamefont{Ashhab}}, \bibnamefont{and}
  \bibinfo{author}{\bibfnamefont{F.}~\bibnamefont{Nori}},
  \bibinfo{journal}{Phys. Rev. A} \textbf{\bibinfo{volume}{82}},
  \bibinfo{pages}{052311} (\bibinfo{year}{2010}).

\end{thebibliography}
\end{document}